\PassOptionsToPackage{pdftex}{graphicx}
\documentclass[twocolumn]{jpsj3}
\usepackage{graphicx}
\usepackage{txfonts}
\usepackage{braket} 
\usepackage{color} 
\usepackage[table]{xcolor} 

\graphicspath{{fig/}}

\title{Quantum Generative Model on Bicycle-sharing System and an Application}

\author{Fumio Nemoto\,$^{1}$\thanks{upwork151@gmail.com}, Nobuyuki Koike\,$^{2}$, Daichi Sato\,$^{3}$, Yuuta Kawaai\,$^{2}$, and Masayuki Ohzeki$^{4,5,6}$}
\inst{$^1$Koto, Tokyo 135-0061, Japan \\
$^2$Sendai City Office, Aoba, Sendai 980-0803, Japan \\
$^3$TAKUMI Solutions Limited, Aoba, Sendai 980-0014, Japan \\
$^{4}$Graduate School of Information Sciences, Tohoku University, Aoba, Sendai 980-8564, Japan \\
$^{5}$Department of Physics, Institute of Science Tokyo, Meguro, Tokyo 152-9551, Japan \\
$^{6}$Sigma-i Co., Ltd., Minato, Tokyo 108-0075, Japan }

\abst{
Recently, bicycle-sharing systems have been implemented in numerous cities, becoming integral to daily life. However, a prevalent issue arises when intensive commuting demand leads to bicycle shortages in specific areas and at particular times.
To address this challenge, we employ a novel quantum machine learning model that analyzes time series data by fitting quantum time evolution to observed sequences. This model enables us to capture actual trends in bicycle counts at individual ports and identify correlations between different ports.
Utilizing the trained model, we simulate the impact of proactively adding bicycles to high-demand ports on the overall rental number across the system.
Given that the core of this method lies in a Monte Carlo simulation, it is anticipated to have a wide range of industrial applications.
}

\begin{document}
\maketitle

\section{Introduction}
Quantum machine learning (QML) promises computational advantages beyond the reach of classical machine learning.
Recent advances have introduced quantum counterparts of classical architectures, such as quantum convolutional neural networks (QCNNs) \cite{cong2019quantum}, quantum recurrent neural networks (QRNNs) \cite{bausch2020recurrent}, and quantum generative adversarial networks (QGANs) \cite{huang2021experimental}.
These models share a standard structure: parameterized quantum circuits that mirror and extend the roles of layers in classical neural networks.

A distinctive line of research \cite{horowitz2022quantum} proposed a parameterized quantum circuit for time-series modeling, directly associating quantum time evolution with multi-dimensional data.
This generative approach leverages the intrinsic dynamics of quantum systems to produce probabilistic time series, while remaining structurally simple.
It has been demonstrated to capture complex stochastic processes, such as drift and correlated Brownian motion.
Subsequent work \cite{okumura2024application} applied this framework to financial data, successfully reproducing short-term market trends and correlations with fewer parameters than conventional models like long short-term memory networks (LSTMs) and vector autoregression models (VARs).

While stock prices require discretization for such models, bicycle-sharing data is inherently discrete.
This natural alignment motivates our study: applying the quantum circuit scheme to bicycle-sharing systems.
By doing so, we aim to capture inter-port correlations and exploit the generative power of quantum circuits to address pressing challenges such as bicycle shortages.

Traditional regression models \cite{zhang2016bicycle,wang2016modeling,ashqar2019modeling} offer flexibility and interpretability but fall short in modeling dynamic interdependencies.
Neural network approaches, particularly graph convolutional neural networks (GCNNs) \cite{lin2018predicting}, address spatial correlations effectively, yet they introduce substantial complexity.

Our contribution takes a different path.
We adapt the quantum generative model of \cite{horowitz2022quantum} to non-stationary bicycle-sharing data.
Unlike classical methods, our approach utilizes quantum time evolution to encode correlations and generate realistic demand scenarios naturally.
Through simulation, we further demonstrate how this generative property can provide practical insights into alleviating bicycle shortages.
In this sense, this work highlights the potential of quantum machine learning as a simpler yet powerful alternative to conventional neural networks, opening new opportunities for demand prediction in mobility systems.
Furthermore, we provide a quantitative comparison with classical baselines, seasonal
autoregressive integrated moving average (SARIMA) and LSTM, demonstrating that our quantum model achieves competitive performance with significantly fewer parameters while effectively recovering inter-port correlations.

\section{Method}
\label{chap:Method}

This section describes quantum machine learning for multi-dimensional time series data. The given multi-dimensional time series are discretized and converted into probability transition matrices that are easy to feed to the quantum circuit.
Meanwhile, a parametrized quantum circuit is designed to express a multi-dimensional time series through its evolution. The parameters are determined so that the output of the quantum circuit emulates the given data.
We implement a unique cost function that evaluates the degree of approximation of individual time series and the degree of approximation of the correlation between time series, making it easy for the quantum system to emulate the actual correlation.

\subsection{Time Series Data}
\label{sec:Time Series Data}

\begin{figure}[h]
    \centering
    \includegraphics[width=0.9\linewidth]{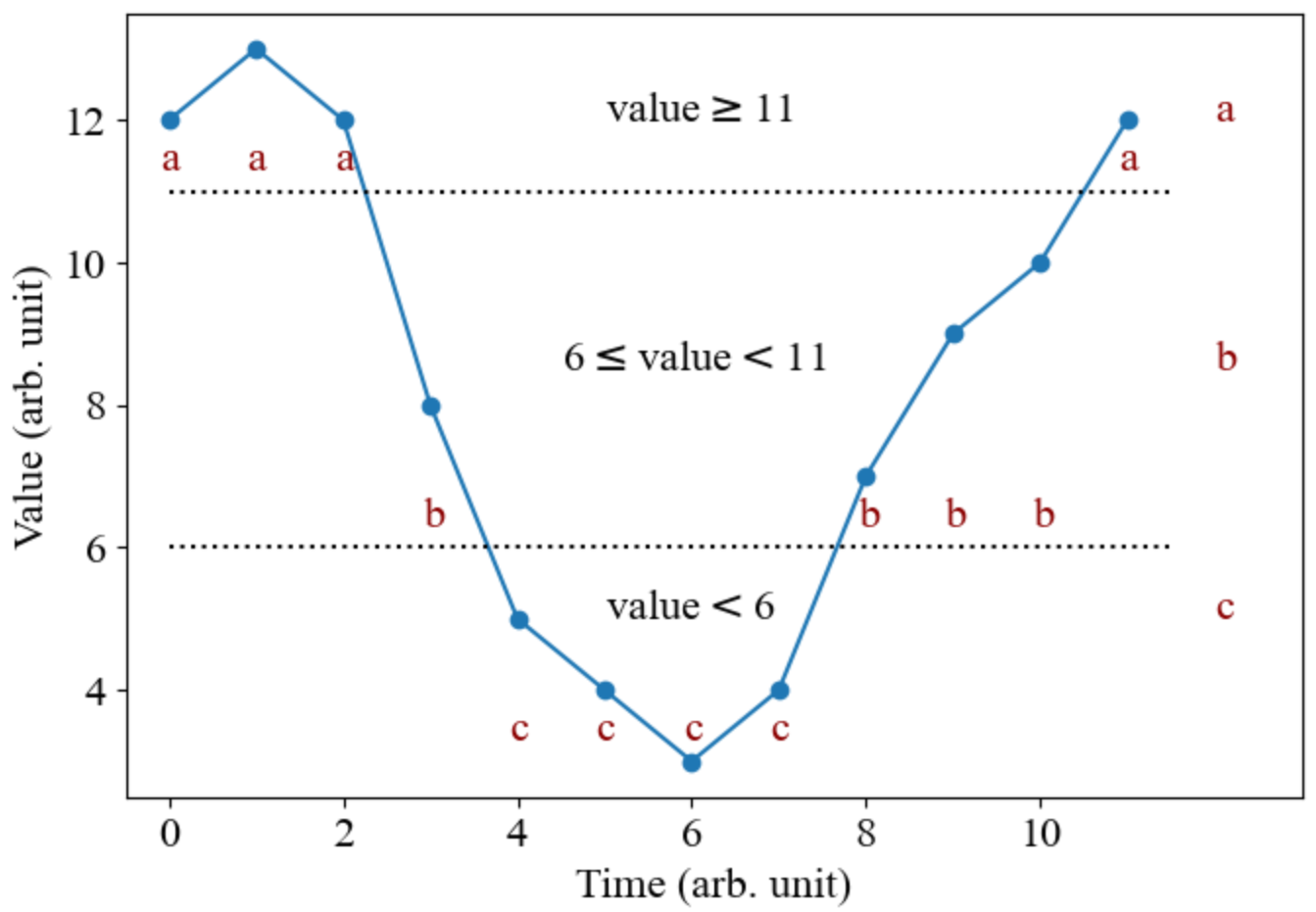}

\caption{(Color online)~Image of SAX. The horizontal axis represents time. In this example, the vertical axis is divided into three intervals (e.g., $y \ge 11$ for `a', $6 \le y < 11$ for `b', $y < 6$ for `c'). By mapping values to the intervals, a time series is converted into a simple character sequence
\{$a,a,a,b,c,c,c,c,b,b,b,a$\}.}
\label{fig:1_1}

\end{figure}

We simplify time series data using the symbolic aggregate approximation (SAX) method \cite{lin2007experiencing}.
We divide the real axis into certain intervals and assign arbitrary symbols such as $a, b, c,\cdots$. A time series is simplified into a set of symbols by converting each data point into a symbol it belongs to. The SAX symbols will map to quantum states later. For this reason, we set the number of intervals, $N$, a power of 2, and call SAX symbols "states". Figure\;\ref{fig:1_1} shows an intuitive image of SAX.
SAX-applied data approximate the original data by adopting the average value of the belongings. We apply the SAX method to each multi-dimensional time series.

\subsection{Probability Transition Matrix}
\label{sec:Probability Transition Matrix}

Since a time series can be transformed into a sequence of $N$ possible states, we can construct an $N \times N$ matrix by aggregating the observed transitions. Specifically, the $(m,n)$ element of the matrix represents the total number of cases in which state $m;(0 \le m < N)$ at time $t-1$ transitions to state $n;(0 \le n < N)$ at time $t$ for any time $t > 0$. The matrix is normalized to obtain a probability-based representation so that the sum of each row equals 1. We refer to this normalized form as the probability transition matrix.
For multi-dimensional time series, a separate probability transition matrix is constructed for each dimension.
We denote $T_{d,t}(n|m)$ as a probability transition matrix for $d$-th time series at time $t$.

\subsection{Quantum Circuit}
\label{sec:quantum circuit}

The quantum circuit consists of target qubits and ancilla qubits. The target qubits represent the states of the time series, while the ancilla qubits enhance the expressiveness of the circuit by enabling entanglement with the target qubits. The number of target qubits assigned to each time series is $\log_2 N$, which generates a subspace whose dimension matches the number of states, $N$.
Accordingly, the time series states ${0, 1, \dots, N-1}$
can be regarded as the quantum orthonormal basis states ${\ket{0}, \ket{1}, \dots, \ket{N-1}}$.
The circuit parameters include a time parameter $t$ and a set of learnable parameters $\vec{\theta}$, which are optimized based on the probability transition matrix.

In our study, the ports of the bicycle-sharing system are assigned to the target qubits. This approach eliminates the need to model individual bicycle trips explicitly and mitigates noise effects, since the probabilistic nature of observations inherently accounts for such variability.

\begin{figure}[h]
    \centering
    \includegraphics[width=1\linewidth]{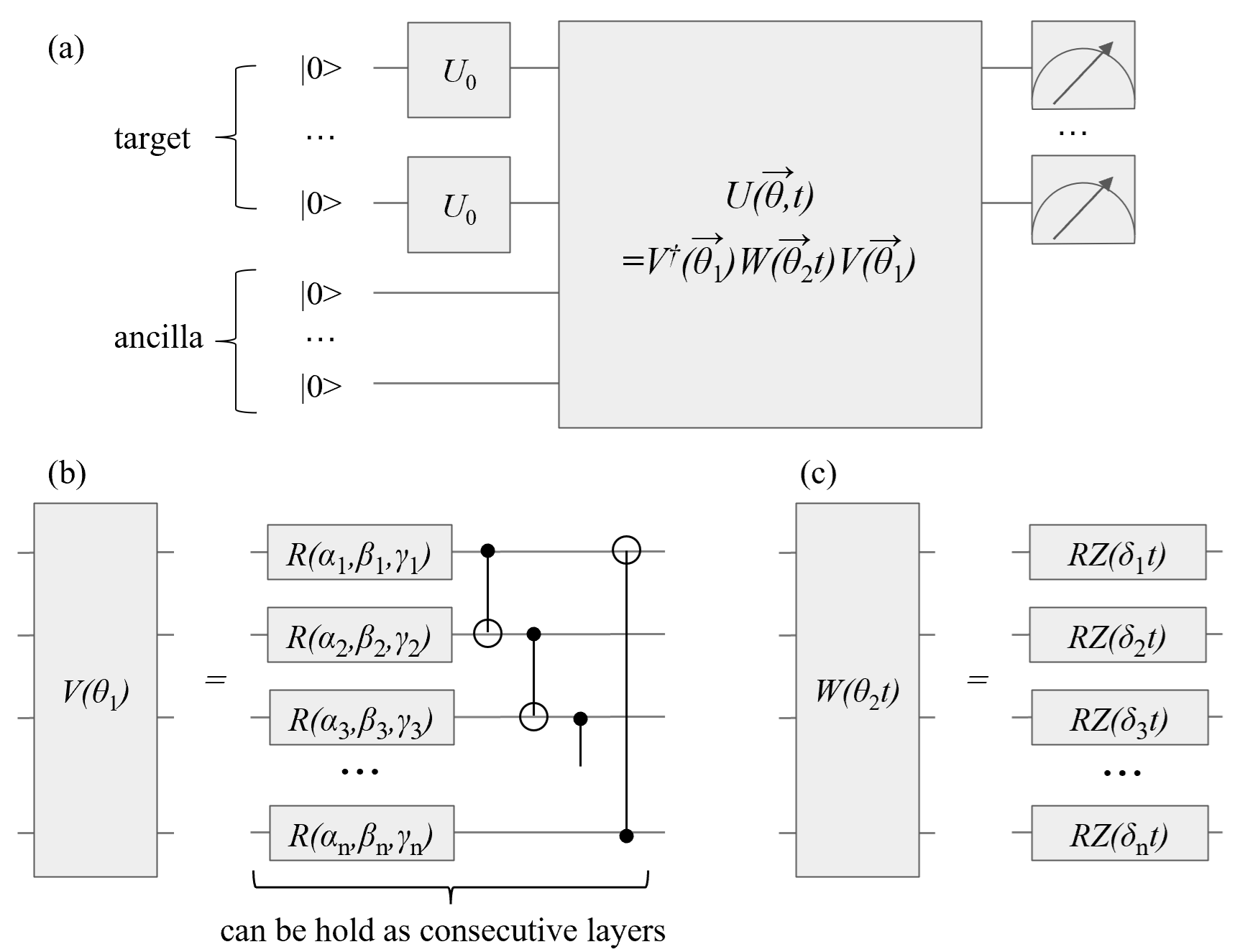}

    \caption{
(a)~Overview of the whole quantum circuit. (b)~Concrete structure of $V$. $R(\alpha_k,\beta_k,\gamma_k)$ can be rewritten as the combination of the three single qubit rotation gates $RZ(\alpha_k)RY(\beta_k)RZ(\gamma_k)$ for each $k\leq{n}$, where $\alpha_k,\beta_k,\gamma_k$ are rotation parameters and $n$ is the total number of qubit in (a).
The $n$ CNOT gates between adjacent qubits are placed after $R$ gates so that every qubit interacts with other qubits, keeping balance of parameter searching and calculation complexity.
We can also build $V$ as another combination pattern of CNOT, e.g., two qubits away, and those different $V$ can be layered \cite{horowitz2022quantum}.
(c) details the orthogonal component $W$. $W$ expresses the time evolution on the eigenspace with searching the eigenvalues $\delta_k$ of Hamiltonian $H$.}

    \label{fig:2_2_1}
\end{figure}

Figure~\ref{fig:2_2_1}~(a) illustrates the overall structure of the quantum circuit.
The operator $U_{0}$ is designed to prepare the initial state of the probability transition matrix at time $0$.
This operator can be constructed by selecting either the Pauli-$X$ gate or the identity operator $I$, depending on the specific initial state.
The operator $U(\vec{\theta},t)$, sometimes referred to as the ans\"atz, represents the time-evolution operator acting on the initial state.
It is defined as
\begin{equation}
    U(\vec{\theta},t) = V^{\dagger}(\vec{\theta_1}) \, W(\vec{\theta_2}t) \, V(\vec{\theta_1}),
\end{equation}
where $\vec{\theta_1}$ and $\vec{\theta_2}$ are reparameterizations of $\vec{\theta}$.
Figure~\ref{fig:2_2_1}~(b) shows the breakdown of the unitary $V(\vec{\theta_1})$, consisting of single-qubit rotation gates followed by CNOT gates which introduce entanglement into the quantum system.
Figure~\ref{fig:2_2_1}~(c) depicts that the operator $W(\vec{\theta_2}t)$ is implemented using $RZ$ rotations on each qubit.
Finally, the final state at time $t$ is obtained by measuring the target qubits.

We intuitively explain $U(\vec{\theta},t)$.
We regard the quantum circuit as an ans\"atz, assuming the existence of some Hamiltonian $H$ such that its time-evolution operator
\begin{equation}
    U = \exp(-\mathrm{i}Ht)
\end{equation}
can reproduce a given probability transition matrix through quantum measurements.

Since $H$ can be diagonalized by a certain unitary operator $V$, we obtain
\begin{equation}
    U = V^{\dagger} \exp(-\mathrm{i}Wt) V,
\end{equation}
where $W$ is a real diagonal matrix.
By transforming the orthonormal basis into the eigenspace of $H$ via $V$, $U$ can be interpreted as a time evolution governed by the diagonal elements of $W$.

Therefore, $U$ is constructed by modeling $V$ and $W$.
The advantage of this decomposition is that it reduces computational cost, since only the diagonal elements of $W$ need to be evaluated when performing calculations for different values of $t$.

\subsection{Mapping Quantum Outputs to Transition Matrices}
\label{sec:Mapping}

The connection between the quantum circuit and the probability transition matrices
is established through measurement.
Given an initial state $m$ at time $0$,
the operator $U_{0}$ prepares the corresponding basis state $\ket{m}$.
The circuit then evolves this state under the parameterized time-evolution operator
$U(\vec{\theta},t)$.
Measuring the target qubits at time $t$ yields outcome $n \in \{0,\dots,N-1\}$ with probability
\[
    P^{\vec{\theta}}_{d,t}(n|m)
    = \bigl|\braket{n | U(\vec{\theta},t) | m}\bigr|^2 .
\]
This probability distribution is directly compared with the $T_{d,t}(n|m)$ obtained from data.
In this way, the quantum circuit acts as a generative model that reproduces the observed transition matrix through repeated sampling.

For multi-dimensional data, the joint measurement outcome
$(n_{0},n_{1},\dots,n_{D-1})$ represents the simultaneous states of all $D$ ports.
The corresponding joint probability $P^{\vec{\theta}}_{d,d',t}(n,n')$
is then used to evaluate correlations across ports in the cost function.

\begin{figure}[h]
    \centering
    \raggedright
    \includegraphics[width=0.9\linewidth]{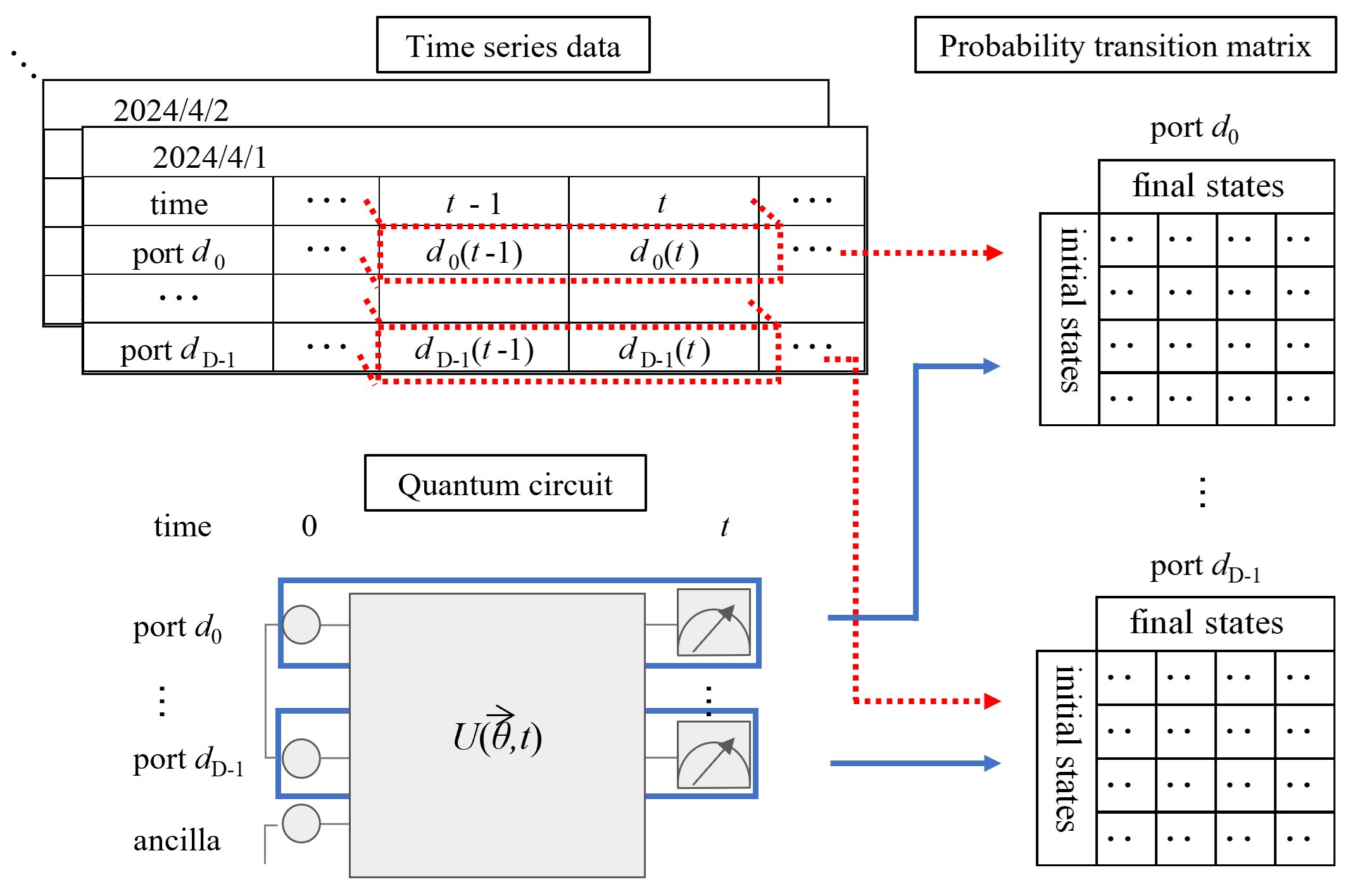}

\caption{(Color online)
Visualizing of the relationship between time series data, transition probability matrix, and quantum circuit. The objects drawn with dotted lines and solid lines represent the aggregation of the probability transition matrices based on real data and the quantum circuit measurements, respectively.}

\label{fig:2_3}
\end{figure}

Figure\;\ref{fig:2_3} summarizes the relationship between time series data, probability transition matrix, and quantum circuit.
This picture focuses on some fixed $t$ but the same relationship holds for any $t$.
The probability transition matrices are aggregated for each port $d$ and time $t$ via SAX discritization by collecting time series data of the same data point. In the case of bicycle-sharing, each probability transition matrix is aggregated by collecting over multiple days data for the same port and time.
Probability transition matrices are also evaluated by measurement of the quantum circuit, which will trained so that these two estimates are close.

\subsection{Cost Function}
\label{Cost Function}

In quantum machine learning, the parameters $\vec{\theta}$ are optimized so that the time evolution of the quantum circuit aligns with the probability transition matrix.
The procedure is outlined as follows.

Both $P^{\vec{\theta}}_{d,t}(n|m)$ and $T_{d,t}(n|m)$ define probability distributions over the $N$ possible states.
Hence, the Kullback--Leibler (KL) divergence between them is expressed as
\begin{equation}
    D_{\mathrm{KL}}\!\left(P^{\vec{\theta}}_{d,t} \;\|\; T_{d,t}\right)
    = \sum_{n} P^{\vec{\theta}}_{d,t}(n|m)
      \log \frac{P^{\vec{\theta}}_{d,t}(n|m)}{T_{d,t}(n|m)}.
      \label{eq:KLinfo}
\end{equation}
By summing over all ports $d$, times $t$, and initial states $m$,
we obtain the overall discrepancy between the model and the probability transition matrices:
\begin{equation}
    \mathrm{L}(\vec{\theta})
    = \sum_{d,t,m} D_{\mathrm{KL}}\!\left(P^{\vec{\theta}}_{d,t} \;\|\; T_{d,t}\right).
\end{equation}
This loss function $\mathrm{L}(\vec{\theta})$ serves as the cost function of our model,
quantifying the divergence between the predicted and observed transition distributions.

However, in general, the absence of explicit correlation information between time series raises the concern that reflecting the correlation structure in the model will be difficult.
In the context of bicycle-sharing systems, the variance and covariance structure across ports play a crucial role in capturing demand fluctuations and inter-port dependencies.
To address this, we introduce additional terms into our quantum model so that these statistical correlations observed in real data can be appropriately incorporated, thereby enabling the model to better reflect the underlying dynamics of the system.

Thus, as a unique idea for our study compared to previous studies, we added a term that evaluated the difference between the correlation coefficients obtained from the quantum model and the actual data. The cost function in our study is as follows.

\begin{equation}
\mathrm{C}(\vec{\theta})=\frac{1}{GN}\mathrm{L}(\vec{\theta})+\frac{1}{G}\sum_{d{\neq}d'}\sum_{t}{\alpha}_{d,d'}{({\rho}_{d,d',t}-{\rho}^{\vec{\theta}}_{d,d',t})}^2,
\label{eq:cost_func}
\end{equation}
where the coverage of the summation of the outer sigma symbol in the second term is any distinct port pair. $G$ is the number of grids in valuable $t$.

The division by constants of both terms intends scale normalization so that the scale of the first term is KL divergence per port multiplied by the number of ports, and the scale of the second term is the squared error per port pair multiplied by the number of port pairs.

$\rho_{d,d',t}$ represents the correlation coefficient in the actual data. The concrete formulation is as follows.
\begin{equation}
\rho_{d,d',t}=\displaystyle\frac{\sum_{x\in{X}}(x_{d,t}-M_{d,t})(x_{d',t}-M_{d',t})}{{\left(\sum_{x\in{X}}{(x_{d,t}-M_{d,t})}^2\right)}^{\frac{1}{2}}{\left(\sum_{x\in{X}}{(x_{d',t}-M_{d',t})}^2\right)}^\frac{1}{2}},
\label{eq:actual_corr}
\end{equation}
where $x\in{X}$ means that $x$ is a multi-dimensional bicycle-sharing data of a day among all data $X$ collected over multiple days.
$x_{d,t}$ is the data point of $x$ at a port $d$ and at time $t$.
$M_{d,t}$ indicates the mean value of $x_{d,t}$ where $\{x~|~x\in{X}\}$.

On the other hand, $\rho^{\vec{\theta}}_{d,d',t}$ represents the correlation coefficient in the quantum model whose formulation is

\begin{equation}
\rho^{\vec{\theta}}_{d,d',t}=\displaystyle\frac{\sum_{n,n'}P^{\vec{\theta}}_{d,d',t}(n,n')(A_{d,t}(n)-M_{d,t})(A_{d',t}(n')-M_{d',t})}{{\left(\sum_{n}P^{\vec{\theta}}_{d,t}(n){(A_{d,t}(n)-M_{d,t})}^2\right)}^{\frac{1}{2}}{\left(\sum_{n'}P^{\vec{\theta}}_{d',t}(n'){(A_{d',t}(n')-M_{d',t})}^2\right)}^\frac{1}{2}},
\label{eq:quantum_corr}
\end{equation}
where $A_{d,t}(n)$ represents the average value of $\{x_{d,t}~|~x\in{X}\}$
under the condition
of state $n$, that is mentioned in Section \ref{sec:Time Series Data}. $P^{\vec{\theta}}_{d,d',t}(n,n')$ represents the joint probability that the states of ports $d$ and $d'$ at time $t$ are $\ket{n}$ and $\ket{n'}$, while $P^{\vec{\theta}}_{d,t}(n)$ and $P^{\vec{\theta}}_{d',t}(n')$ denote their respective marginal probabilities.

The second term in Eq.~\ref{eq:cost_func} evaluates squared errors for every time grid $t$. However, in implementation, we can set time granularity to evaluate squared errors arbitrary depending on the purpose. The finer the evaluation granularity of the second term, the higher the complexity of the model required.

$\alpha_{d,d'}$ is a hyperparameter that controls the prioritization of reducing the cost function related to individual port pairs. The effect of tuning $\alpha_{d,d'}$ is explained in Section~\ref{Section:Prediction and Validation}.

\subsection{Multiple Ports and Qubit Representation}
\label{sec:multiports}

In our experiment, we use time series of one-hour increment of bicycles rather than raw count to more directly reflect demand pressure.
The number of states $N$ is set to 2 per dimension, which means each time-grid increment
$\Delta x_{d,t}$ is discretized into two SAX states (low/negative vs.~high/positive).
The measurement outcome $n_{d,t}\in\{0,1\}$ is mapped back to a representative increment
$A_{d,t}(n_{d,t})=\mathbb{E}[\Delta x_{d,t}\mid \text{state }n_{d,t}\in\{0,1\}]$ estimated from data.
While higher $N$ (e.g., 4 or 8) could capture finer granularity, $N=2$ was chosen for computational efficiency and because the binary distinction between "decreasing (shortage risk)" and "increasing (return flow)" is the dominant factor for demand rebalancing simulations.
Starting from the initial count $x_{d,0}$, the daily trajectory is reconstructed by
\begin{equation}
    x_{d,t+1}=x_{d,t}+A_{d,t}(n_{d,t}).\label{stoc}
\end{equation}

Although each port is represented by only two SAX states ($N=2$),
the actual number of bicycles can fluctuate by several hundreds.
This apparent gap is bridged by the mapping step and the accumulation over time.
By sequentially sampling the circuit and accumulating these increments,
we obtain full-day trajectories of bicycle counts.
Repeating this procedure many times (e.g., $1000$ sample paths)
allows us to capture realistic, large-scale fluctuations as seen in Fig.~\ref{fig:4_3} and Fig.~\ref{fig:5_3}.

Each port (or port group) is treated as one dimension of the
multi-dimensional time series, and thus requires $\log_2 N = 1$ target qubit.
For example, for three ports (or port groups) $\mathrm{P_1},\mathrm{P_2},\mathrm{P_3}$, we assign three 
target qubits in the same circuit.

A joint measurement at time $t$ produces a bit string
$(n_{\mathrm{P_1}}, n_{\mathrm{P_2}}, n_{\mathrm{P_3}}) \in \{0,1\}^3$,
which is mapped back to representative increments of bicycle counts for the
corresponding ports.
The parameterized circuit includes entangling gates among these target qubits
and ancilla qubits enabling the model to learn both intra-port dynamics
and inter-port correlations simultaneously.
The additional correlation term in the cost function further enforces this property.

All quantum circuits in this study were executed on a classical simulator
(shot-free), using \texttt{PennyLane}.
Each circuit evaluation used 25,600 times analytic probability calculation.
Unless otherwise stated, no hardware noise model was applied.

The circuit complexity metrics are summarized as follows. The total number of qubit is six, consisting of three target qubits and three ancillary qubits. All entangling gates in the circuit are CNOT gates in $V$.
We set the unitary operator $U$ with three layers of $V$.
As a result, the circuit depth is estimated to be 44. The number of two-qubit gates is 36, all of which are CNOT gates~(see Fig.~\ref{fig:2_2_1} for the quantum curcuit diagram).

\section{Numerical Experiments and Simulation}
\label{chap:Numerical Experiment}

This section applies the method described in Section~\ref{chap:Method} to a real bicycle-sharing system and then uses the trained quantum generative model for a counterfactual simulation.
Concretely, we (i) prepare multi-dimensional time series via SAX discretization (Section~\ref{sec:Time Series Data}),
(ii) construct probability transition matrices and map them to the quantum circuit outputs as in Section~\ref{sec:Mapping} (see also the measurement-to-transition correspondence introduced between Eqs.~(1)--(4) and Section~\ref{Cost Function}),
(iii) learn circuit parameters by minimizing the cost in Eq.~(\ref{eq:cost_func}),
and (iv) validate both marginal dynamics and cross-port correlations using Eqs.~(\ref{eq:actual_corr}) and~(\ref{eq:quantum_corr}).
Finally, we exploit the model's generative nature to estimate the effect of pre-adding bicycles.

\subsection{Data Preparation}
\label{chap:Data Preparation}

\vskip\baselineskip

The bicycle-sharing system targeted in our study is DATE BIKE in Sendai City.
DATE BIKE covers the central area of Sendai, with 134 operational ports as of April 2024.

A data overview before machine learning showed that the bicycle-sharing system is strongly affected by commuter use. As a whole system, bicycles in residential areas tended to move to office areas in the morning and return in the evening.
Thus, we classify all 134 ports into three groups: "\textit{Residential}," "\textit{Office}," and "\textit{Others}." \textit{Residential} includes all ports where the average bicycle count decreases by two or more from 7:00 to 9:00 on weekdays, \textit{Office} consists of all ports where the average bicycle count increases by two or more from 7:00 to 9:00 on weekdays, and \textit{Others} includes all other ports.
The total number of bicycles within each group is then aggregated, treating the system as effectively consisting of three port groups.
Table \ref{table:Port Aggregation} summarizes the classification.
See Appendix \ref{App:Port Classification} for details that the grouping consistently reflects the current Sendai City.

\begin{table}
\begin{center}
\begin{tabular}{lll}
\hline
\multicolumn{1}{c}{Group} & \multicolumn{1}{c}{Number of Ports}& \multicolumn{1}{c}{Number of Racks} \\
\hline
\textit{Residential} & 50(37.3\%) &\ 453(35.5\%) \\
\textit{Office} & 36(26.8\%) &\ 391(30.7\%) \\
\textit{Others} & 48(35.9\%) &\ 431(33.8\%) \\
\hline
\hline
Total & 134(100\%) &\ 1275(100\%) \\
\hline
\end{tabular}
\end{center}
\caption{Aggregation of ports by group.}
\label{table:Port Aggregation}

\end{table}

Following Section~\ref{sec:Time Series Data}, we discretize the one-hour increment $\Delta x_{d,t}$ at each group $d$ and time grid $t$ using SAX with $N=2$ states.
We fix the start of day at 6{:}00  and set $t=0$ there; then $t$ increases by one per hour until 22{:}00 ($t=16$).
This fixed origin is chosen because the demand is non-stationary over the day, unlike stationary settings in Refs.~\cite{horowitz2022quantum,okumura2024application}.

The data period is April 1–30, 2024; we use 21 weekdays, thus preparing 21 realizations of a 3-dimensional time series on a 17-point time grid.

\subsection{Learning via Cost Function Minimization}
\label{chap:Learning via Cost Function}

For each port group $d$, we aggregate empirical transitions to build $T_{d,t}(n|m)$ (row-normalized) as in Section~\ref{sec:Probability Transition Matrix}.
The quantum circuit of Section~\ref{sec:quantum circuit} prepares the initial basis $\ket{m}$ using $U_{0}$, evolves it by $U(\vec{\theta},t)$, and yields the final outcome $n$ with probability $P_{d,t}^{\vec{\theta}}(n|m)$ upon measurement.

We estimate $P_{d,t}^{\vec{\theta}}(n|m)$ by a \texttt{PennyLane}’s function that computes exact probabilities, and then optimize $\vec{\theta}$ by minimizing the composite objective in Eq.~(\ref{eq:cost_func}), i.e., the sum of the KL terms comparing $P_{d,t}^{\vec{\theta}}(\cdot|m)$ and $T_{d,t}(\cdot|m)$, plus the correlation penalty based on Eqs.~(\ref{eq:actual_corr})--(\ref{eq:quantum_corr}).
This procedure enforces consistency at both the marginal (per-port) and joint (cross-port) levels.

We evaluate the correlation term in Eq.~(\ref{eq:cost_func}) at a time granularity of three periods: 6{:}00 - 11{:}00, 11{:}00 - 16{:}00, and 16{:}00 - 22{:}00. This is intended to divide the time into morning and evening that are affected by commuting work and home, and afternoon that are not. It makes sense to build a model that recovers correlation time structure which reflect the daily commuting pattern for use in a simulation of the commuting issue in Section \ref{chap:Simulation}.

Hyperparameters are as follows: the number of SAX states is $N=2$, so each transition matrix is $2{\times}2$ and indices $m,n\in\{0,1\}$ in Eq.~(\ref{eq:cost_func}).
We set ${\alpha}_{d,d'}~\equiv~1$ for all port group pair $d,d'$, use Adam~\cite{kingma2014adam} with learning rate $0.1$ for $200$ iterations, and estimate probabilities by analytical calculation of state vector on a classical simulator (see Section \ref{sec:multiports} for details on mapping from measured states to increments).

\begin{figure}[h]
    \centering
    \includegraphics[width=1.0\linewidth]{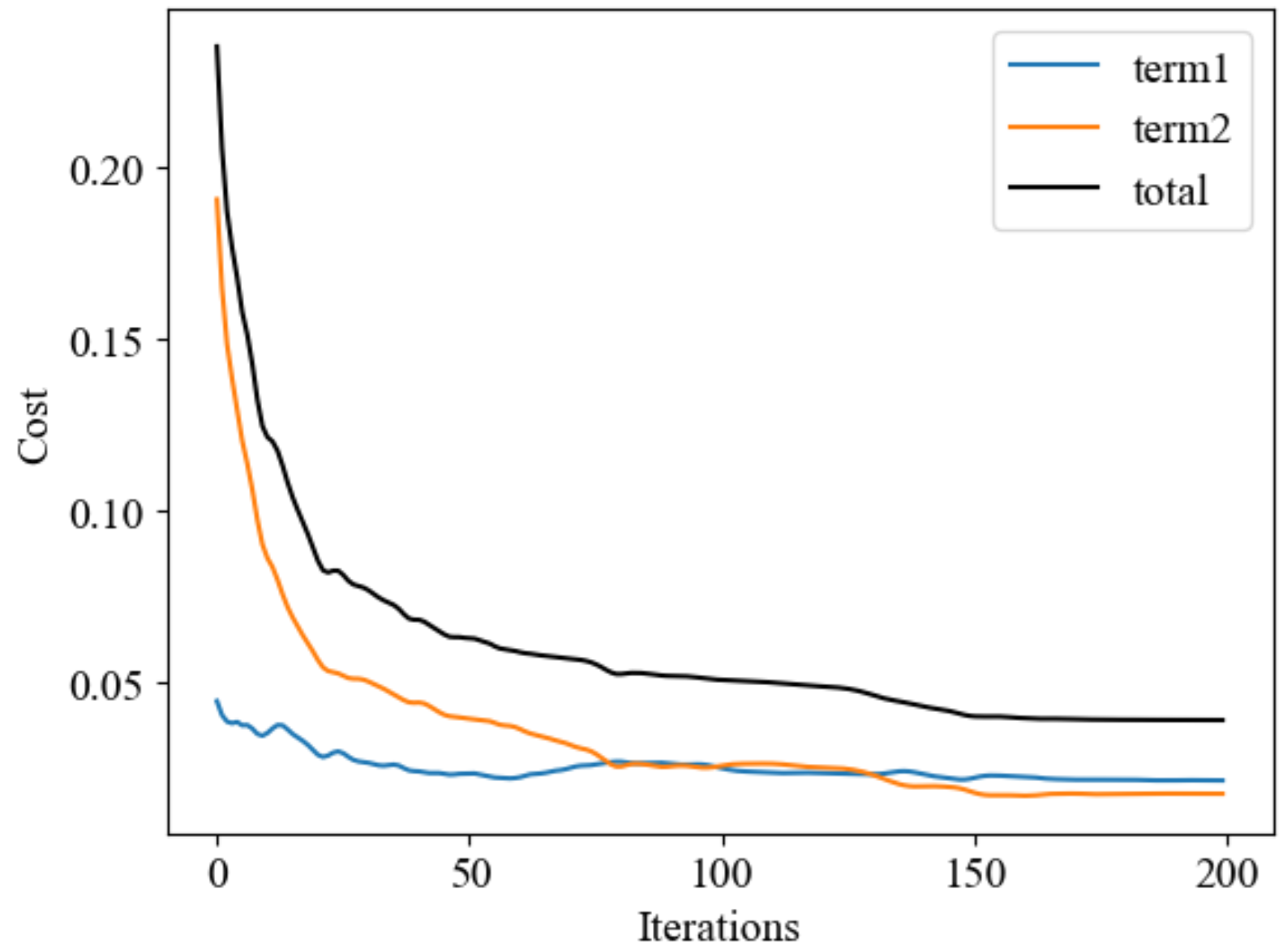}
    
\caption{(Color online)~Cost change over the 200 iteration. "term1", "term2" shows the first and second term in formula (\ref{eq:cost_func}).
The result shows that the cost-minimizing process is sufficiently converged.}
\label{fig:4_6}

\end{figure}

Figure~\ref{fig:4_6} reports the decrease and stabilization of each term in Eq.~(\ref{eq:cost_func}) and the total.
Setting initial parameters randomly commonly results in little correlation among qubit pairs, so the cost in the second term of Eq.~(\ref{eq:cost_func}) start from small value for uncorrelated port pairs but large value for correlated (or negative correlated) port pairs. When it comes to our bicycle-sharing data, \textit{Residential}~/~\textit{Office} and \textit{Others}~/~\textit{Residential} pair, those have negative correlation coefficient in the morning and evening, boosted up the initial cost in the the second term and took long steps to converge.

\subsection{Prediction and Validation of Dynamics and Correlations}
\label{Section:Prediction and Validation}

\begin{figure}[h]
    \centering
    \raggedright
    \includegraphics[width=0.95\linewidth]{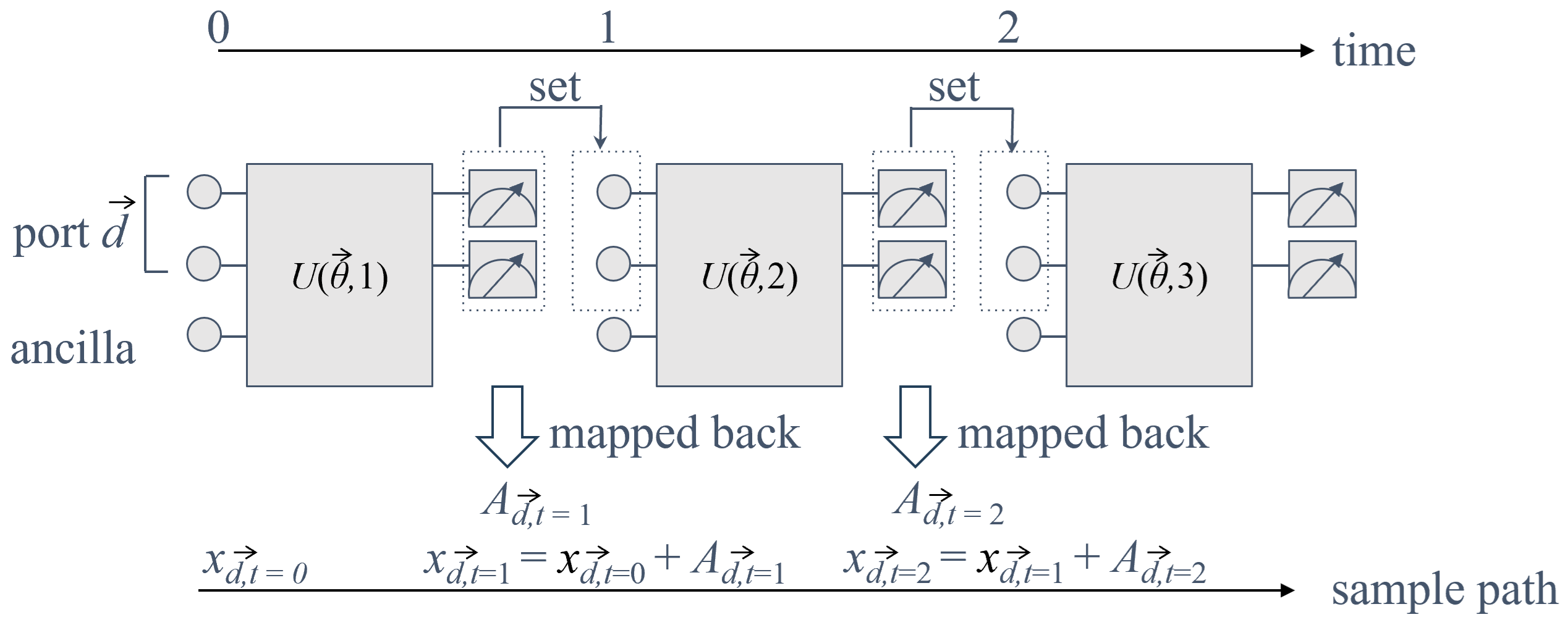}
\caption{
In this figure, $\vec{d}$ denotes a multi-dimensional port. The iterative processes of time evolution, measurement, and mapping back are performed simultaneously on multiple ports, resulting in a multi-dimensional sample path.}
\label{fig:4_2}
\end{figure}

Given the trained circuit, we generate multi-dimensional sample paths by iterating $t=1\to16$ and measuring all target qubits simultaneously at each $t$.
Each outcome $n_{d,t}\in\{0,1\}$ is mapped back to the representative increment $A_{d,t}(n_{d,t})$, and daily trajectories are reconstructed by running Eq. (\ref{stoc}).
We repeat this procedure to obtain many sample paths and use their averages or distributions for validation. Figure \ref{fig:4_2} sketched how to obtain a sample path.


We use all the prepared 21 weekdays data as training data, and verified whether the sample paths generated by trained model capture the characteristics of real data by comparing outputs of the model with the training data itself.

\begin{figure}
    \centering
    \begin{minipage}{0.8\linewidth}
        \centering
        \includegraphics[width=\linewidth]{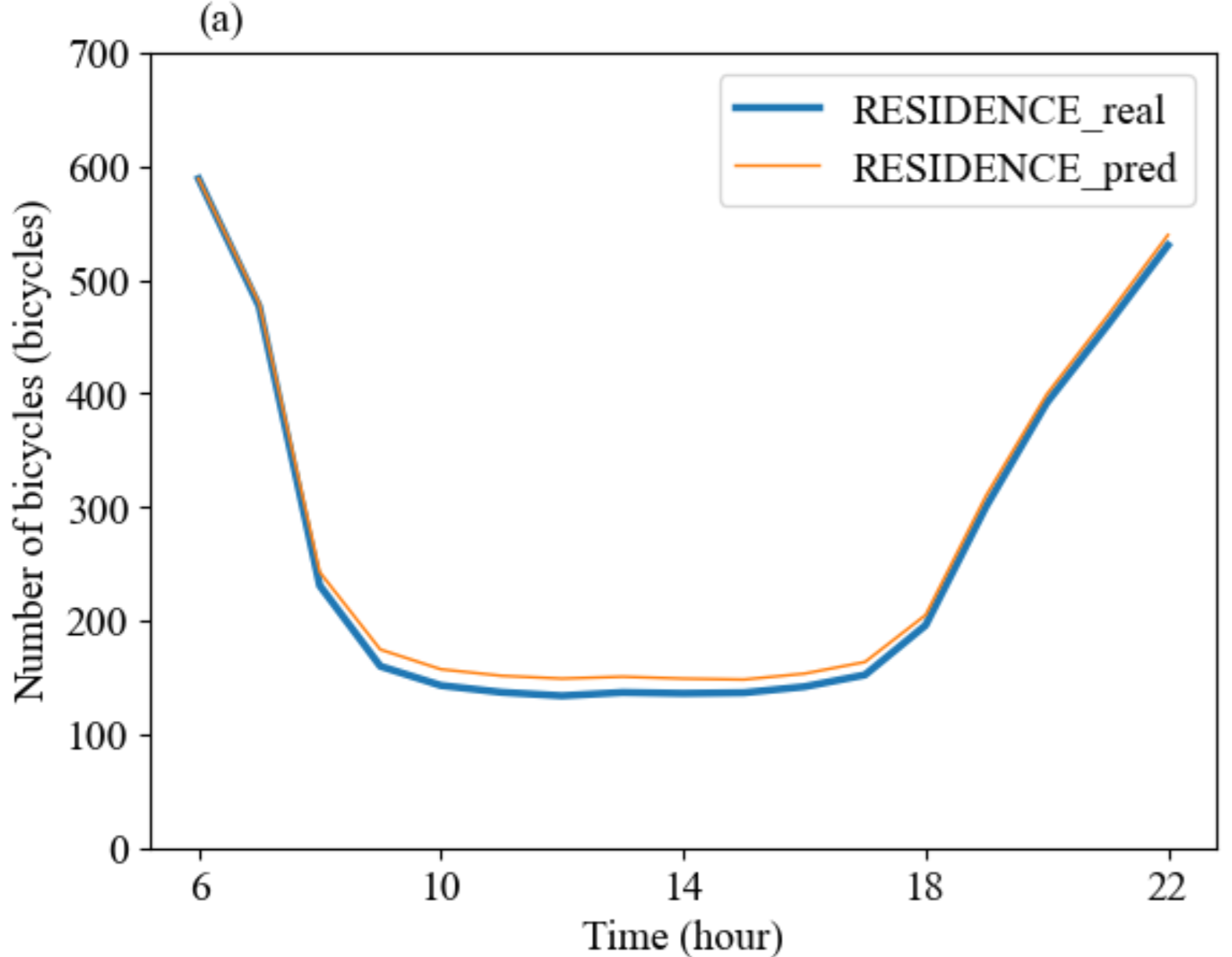}
    \end{minipage}
    \hspace{0.02\linewidth} 
    \begin{minipage}{0.8\linewidth}
        \centering
        \includegraphics[width=\linewidth]{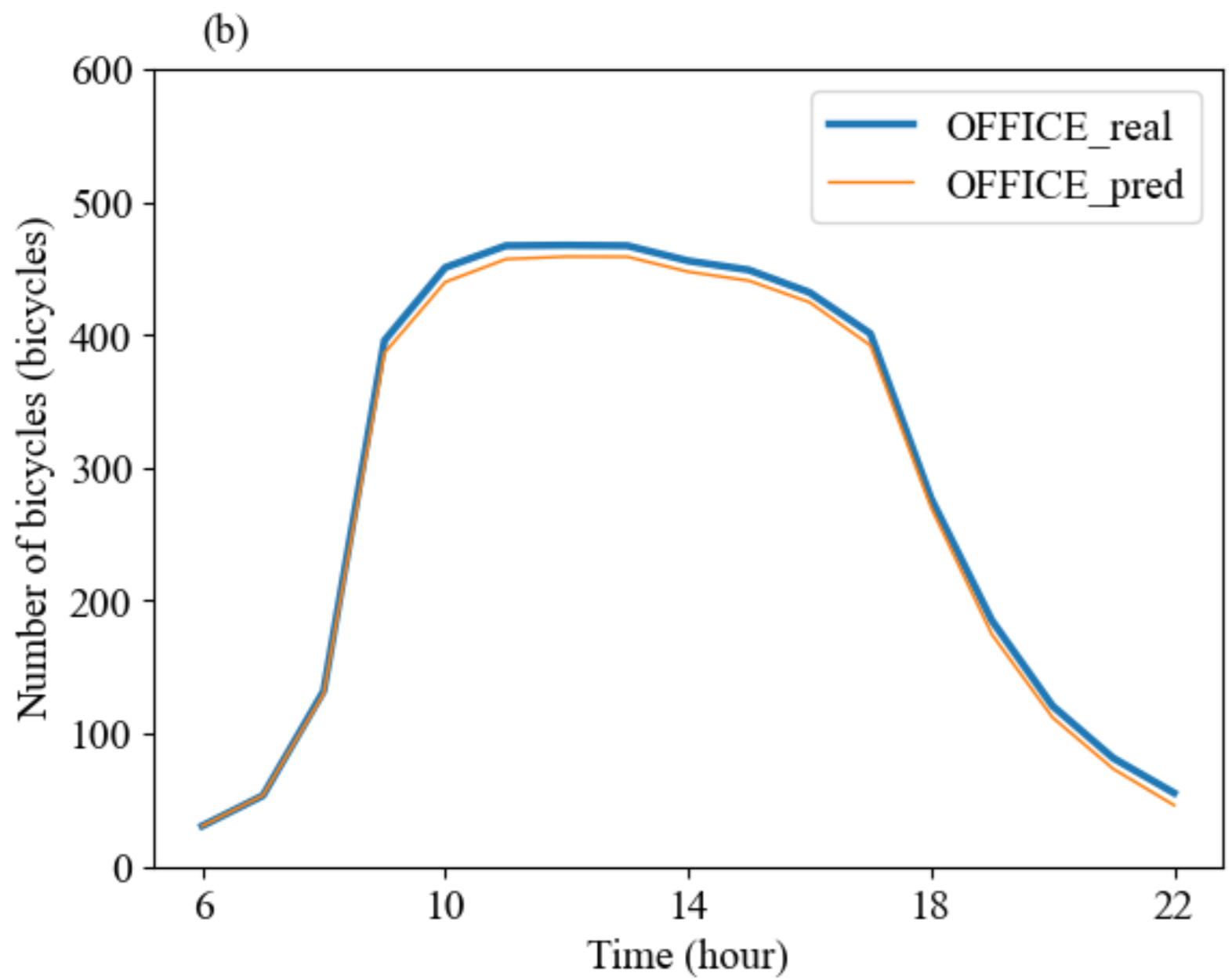}
    \end{minipage}

    \vspace{0.03\linewidth} 
    \begin{minipage}{0.8\linewidth}
        \centering
        \includegraphics[width=\linewidth]{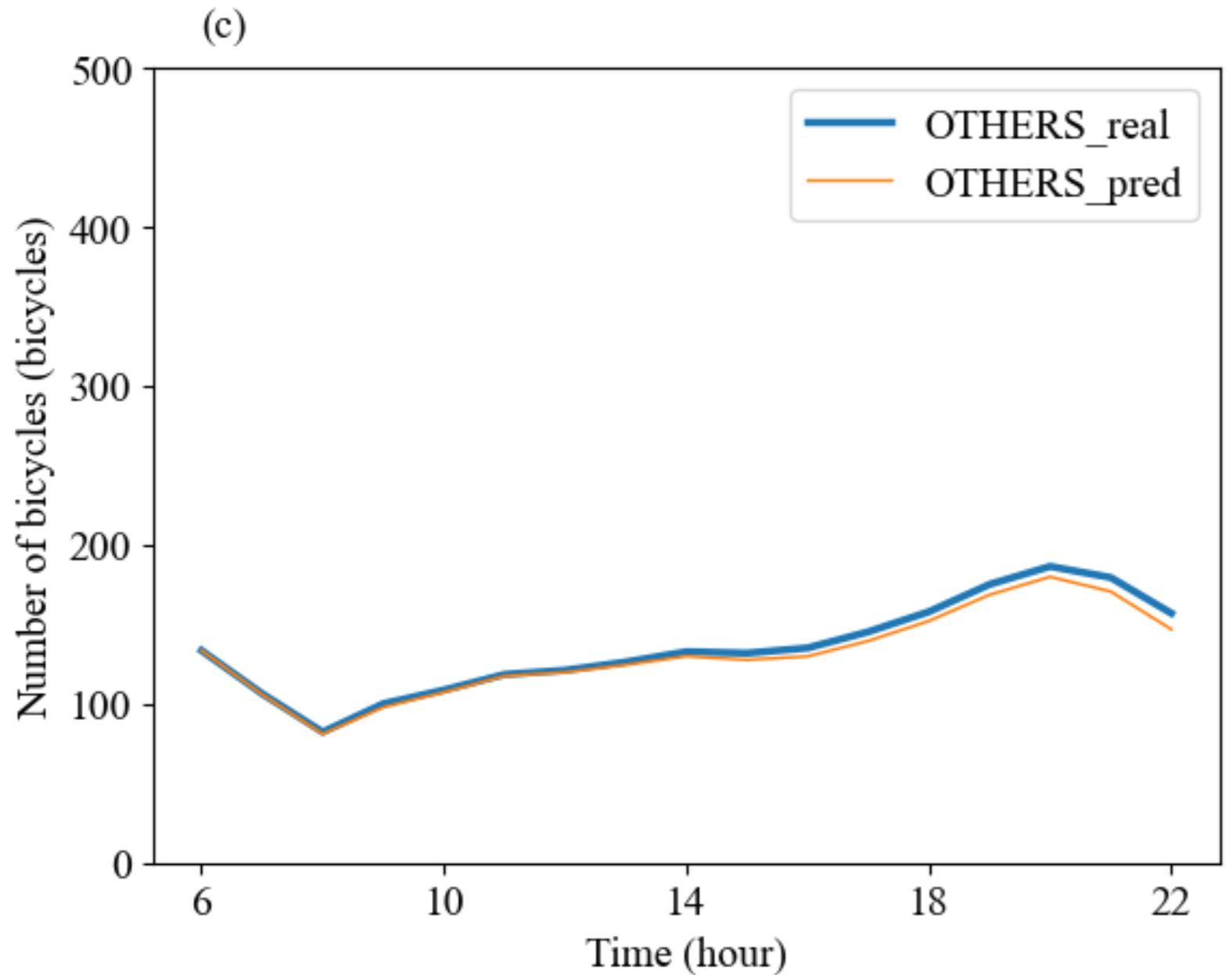}
    \end{minipage}
\caption{
(Color online)~Comparing the absolute bicycle count aggregated by the port group with actual data and sample paths.
The blue (bold) line represents the average daily bicycle count curve for 21-weekdays in April 2024, while the orange line represents the average of 1000 sample paths. Note that the bicycle count in this study is based on the counting available bicycles at ports, thus bicycles on trips are not counted at any port. For this reason, the total number of bicycles in the system is not be constant throughout the day.}
\label{fig:4_3}
\end{figure}

Figure~\ref{fig:4_3} compares the average of 1000 quantum-generated sample paths (orange) against the average of 21-day actual data (blue) for each group.
The trained model captures the characteristic daily trends across all groups, consistent with the KL-fitting term in Eq.~(\ref{eq:cost_func}).
We quantify this fitness by mean absolute error~(MAE) and list the result, $MAE=4.8$, in Table \ref{table:correlation coefficient comparison}. In our study, MAE means the absolute error between sample paths and actual data normalized per port and time, formulated as follows.
\begin{equation}
MAE=\frac{1}{DG}\Sigma_{d,t}|{\bar{x}^m(d,t)}-{\bar{x}^a(d,t)}|,
\end{equation}
where $\bar{x}^m(d,t)$ and $\bar{x}^a(d,t)$ are the average values of sample paths and actual time-series data at data point of port $d$ and time $t$. $D$ is the total number of port groups and $G$ is the total number of time grids.

We use the deviations of hourly increments and evaluate empirical correlations to assess whether inter-port dependencies are reproduced.

Table \ref{table:correlation coefficient comparison} summarize the comparison of correlation coefficient between model and actual. In this table, the column with the label ``$\alpha\equiv1$'' shows the model output under the setting $\alpha\equiv1$, demonstrating that the quantum-generated data recover the qualitative patterns of the real data  (negative \textit{Residential}~/~\textit{Office} and \textit{Others}~/~\textit{Residential} correlation due to commuting flow and weak correlations in the afternoon), aligning with the correlation term in Eq.~(\ref{eq:cost_func}) computed via Eqs.~(\ref{eq:actual_corr}) and~(\ref{eq:quantum_corr}).
It is noteworthy that the model recovers the time-dependent correlation structure.

The cost curve in Fig.~\ref{fig:4_6} demonstrates sufficient convergence, but whether the trained model is stable needs to be verified. The verification is shown in Appendix \ref{App:Model Stability}.

Here, we explain the effect of tuning $\alpha$ in second term of Eq.~(\ref{eq:cost_func}). As referred in Section~\ref{chap:Learning via Cost Function}, when minimizing the cost function Eq.~(\ref{eq:cost_func}), it will take longer steps to reduce cost for those port pairs that have larger absolute value of correlations. This tends to result that such port pairs do not reduce their cost and are worse at reproducing correlation relatively.
To correct this imbalance, we can set some value of $\alpha > 1$ for port pairs with large absolute values of correlation.
Table \ref{table:correlation coefficient comparison} presents the results of setting $(\alpha_\mathrm{Res/Off},\alpha_\mathrm{Off/Oth},\alpha_\mathrm{Oth/Res}) = (20, 1, 20)$, along with $\alpha \equiv 1$, showing that we have better correlation reproduction between \textit{Residential}~/~\textit{Office} and \textit{Others}~/~\textit{Residential} than in case of setting $\alpha \equiv 1$. However, note that MAE tend to be worse as the trade-off for good fitness of correlations.

\begin{table}
\centering
\begin{tabular}{l|ccc}
\hline
 & model & model    & real\\
 & ${\alpha \equiv 1}$ & ${\alpha = (20,1,20)}$    & \\ \hline\hline
MAE & 4.8   & 9.1   &  \text{-} \\ \hline
Correlation Recov & & &\\
~~~~6:00-11:00 & & &\\
 ~~~~~~~~Res / Off & -0.22 & -0.37 & -0.38 \\
 ~~~~~~~~Off / Oth & 0.08  & 0.00  & 0.13 \\
 ~~~~~~~~Oth / Res & -0.11& -0.03   & -0.03 \\
~~~~11:00-16:00 & & &\\
 ~~~~~~~~Res / Off & 0.00& -0.09   & -0.10 \\
 ~~~~~~~~Off / Oth & -0.06& 0.00   & -0.11 \\
 ~~~~~~~~Oth / Res & -0.05& -0.02   & -0.02 \\
~~~~16:00-22:00 & & &\\
 ~~~~~~~~Res / Off & -0.33& -0.43   & -0.44 \\
 ~~~~~~~~Off / Oth & -0.01& 0.03   & -0.02 \\
 ~~~~~~~~Oth / Res & -0.40& -0.48   & -0.49 \\ \hline

\end{tabular}

\caption{Correlation coefficient comparison with model and actual.}
\label{table:correlation coefficient comparison}
\end{table}

\subsection{Performance comparison with other models}
\label{chap:Performance}

In Section~\ref{Section:Prediction and Validation}, we demonstrated that our quantum model successfully reproduces trends and correlations. Here, we compare our model with other typical two models, SARIMA and LSTM, for several performance metrics.

The model parameter of SARIMA is (1,1,1) and (1,1,1,24), which means that the model is based on the ARIMA with all degree of auto-regression, differencing, and moving average are 1, and there are additional parameters to capture 24-hour periodic fluctuations with all degree of auto-regression, differencing, and moving average are also 1.
The LSTM is a multivariate single-layer LSTM model with 50 hidden units and a 3-dimensional dense output layer which learns the next value for the last 24 hours data. Each model was trained on 16-day bicycle count data from April 1 - April 22. After that, trained models output 1000 paths of the bicycle number trajectories from 6:00 to 22:00.
The source of randomness in the paths is random term included in the model for SARIMA, the dropout method \cite{gal2016dropout} for LSTM.

The metrics compared were MAE, reproducibility of correlations, the number of parameters, and training time. MAE and the reproducibility of correlations are compared with 6 days of test data from April 23 - April 30.

Table \ref{table:performance} shows the comparison results.
The quantum model recorded the best score of MAE.
Regarding correlation, the quantum model captures the unique characteristics of bicycle-sharing system, such as the negative correlation between \textit{Residential} and \textit{Office} in the morning and evening, and decrease in correlations across the system in the afternoon.
The correlation between \textit{Office}~/~\textit{Others} and \textit{Others}~/~\textit{Residential} of 6:00 - 11:00 in the actual data (gray cells) were not reproduced well by any of the systems because the distinction from the corresponding training data was very large, over 0.5.

The number of quantum model parameters is 60, which is much fewer than LSTM. SARIMA has the fewest parameters, but it is unsuitable for reproducing correlations since it does not have a mechanism for reflecting spatial correlations.
In terms of training time, the quantum model took the longest. However, it should be noted that a classical simulator was used.
The spent time, 70 minutes, can be broken down into 13 minutes of circuit execution time and the rest of parameter updating time. The time required for one calculation of the state after time evolution was about 0.03 seconds, total 25,600 times calculation read to 13 minutes.
As the quantum circuit for our model is not particularly deep, it can be expected that training will take less time if an actual quantum machine is used.

\begin{table}
\centering

\small 

\begin{tabular}{l|cccc}
\hline
  
 & Quantum & SARIMA & LSTM  & real\\ \hline\hline
 MAE  & 17.1    & 46.2  & 56.8  & \text{-}   \\   \hline
Number of Params  & 60 & 15 & 10,953 & \text{-}  \\  \hline
Training Time (min) & 70 & 1 & 7 & \text{-}  \\ \hline

Correlation Recov & & & & \\
~~~~6:00-11:00 & & & & \\
 ~~~~~~~~Res / Off & -0.30 & 0.00 & -0.26 & -0.29 \\
 ~~~~~~~~Off / Oth & 0.01 & -0.02 & 0.23 & \cellcolor{gray!20}0.50 \\
 ~~~~~~~~Oth / Res & 0.04 & 0.01 & 0.42 & \cellcolor{gray!20}-0.74 \\
~~~~11:00-16:00 & & & & \\
 ~~~~~~~~Res / Off & -0.22 & 0.02 & -0.61 & 0.27 \\
 ~~~~~~~~Off / Oth & 0.00 & 0.00 & -0.08 & -0.10 \\
 ~~~~~~~~Oth / Res & -0.07 & -0.01 & -0.05 & 0.09 \\
~~~~16:00-22:00 & & & & \\
 ~~~~~~~~Res / Off & -0.30 & 0.01 & -0.58 & -0.46 \\
 ~~~~~~~~Off / Oth & -0.09 & 0.01 & -0.32 & -0.32 \\
 ~~~~~~~~Oth / Res & -0.33 & 0.01 & -0.04 & -0.40 \\ \hline

\end{tabular}

\caption{Result of performance comparison with other models.}
\label{table:performance}
\end{table}

\subsection{Simulation: Effect of Pre-Adding Bicycles}
\label{chap:Simulation}

We next use the trained generator for a counterfactual intervention: add bicycles to residential ports at 6{:}00 and estimate the resulting increase in rentals (``effect'').
Formally, letting $X^a_t$ denote the bicycle-count process under addition $a$ and $x^a(t)$ a sample path, the daily effect is defined by
$\max(a-\min_t x^a(t),0)$, which is path-dependent and thus well-suited to Monte Carlo with our generative model.

Since the bicycle-sharing data are based on actual rentals, they do not directly contain information on opportunity losses.
Therefore, we estimate rental opportunity losses under certain assumptions and modify actual bicycle count data to reflect rental opportunity losses. The detail on estimating opportunity losses is in Appendix\ref{App:Estimating Opportunity Losses}.

To conduct a simulation, we train a quantum circuit and generate 1000 sample paths from 6{:}00 to 22{:}00.
These are the sample paths without bicycle addition. A classical computer computes the effect of adding 100 bicycles to the ports in the residential area at 6:00.

Consider a situation where we added bicycles to port A and got one effect at A, i.e., an additional rental from A to some port B occurred. This means that an additional bicycle is supplied to B compared to the case of no addition to A. This suggests that another effect can arise at B if there are enough rentals so all B bicycles run out.
We denote the effect at the port where bicycles are added initially as the "primary effect" and the effect at the port where the destination of an additional bicycle trip caused by the primary effect as the "secondary effect."
The simulation measures up to the secondary effect.

\begin{figure}
    \centering
    \begin{minipage}{0.8\linewidth}
        \centering
        \includegraphics[width=\linewidth]{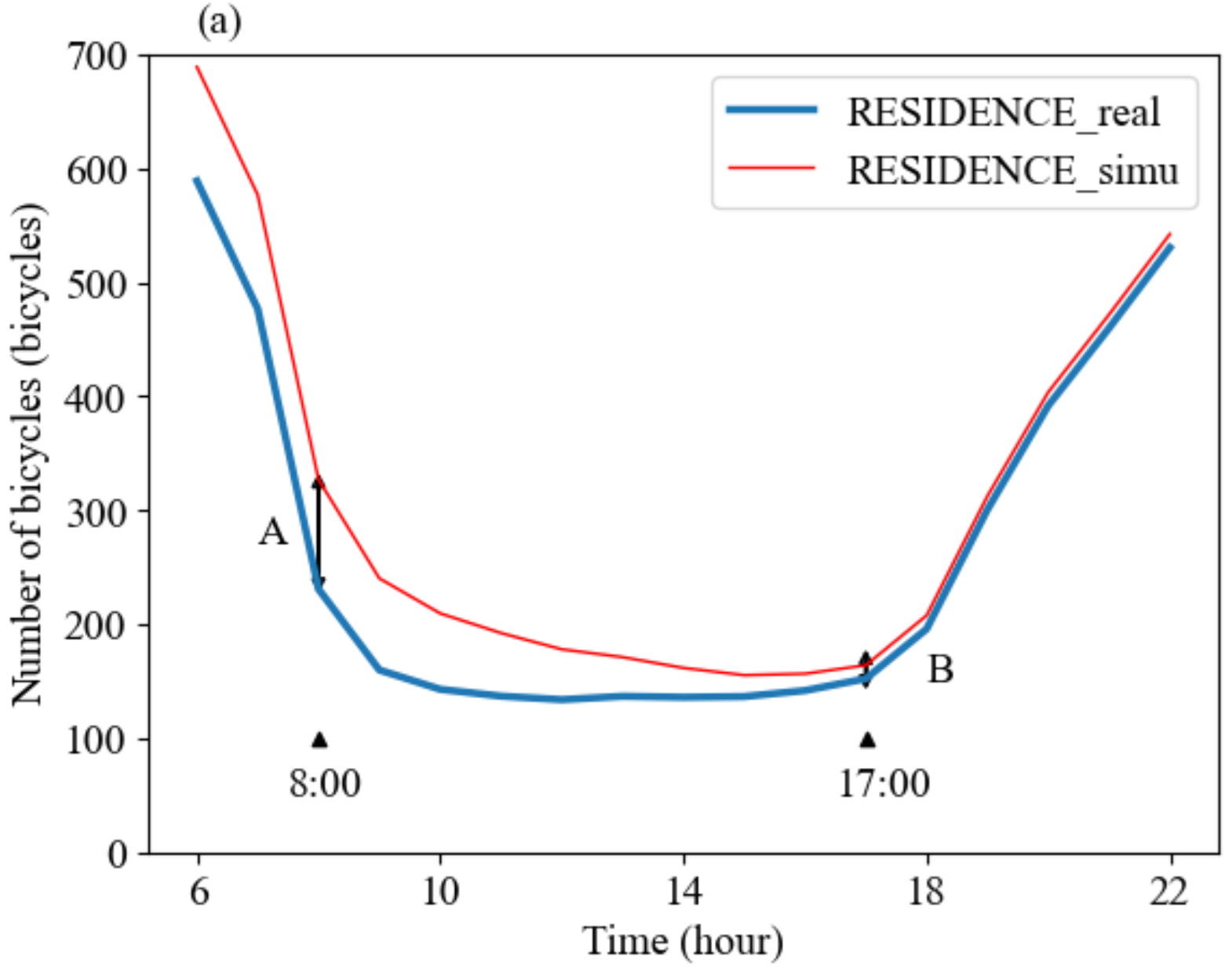}
    \end{minipage}
    \hspace{0.02\linewidth} 
    \begin{minipage}{0.8\linewidth}
        \centering
        \includegraphics[width=\linewidth]{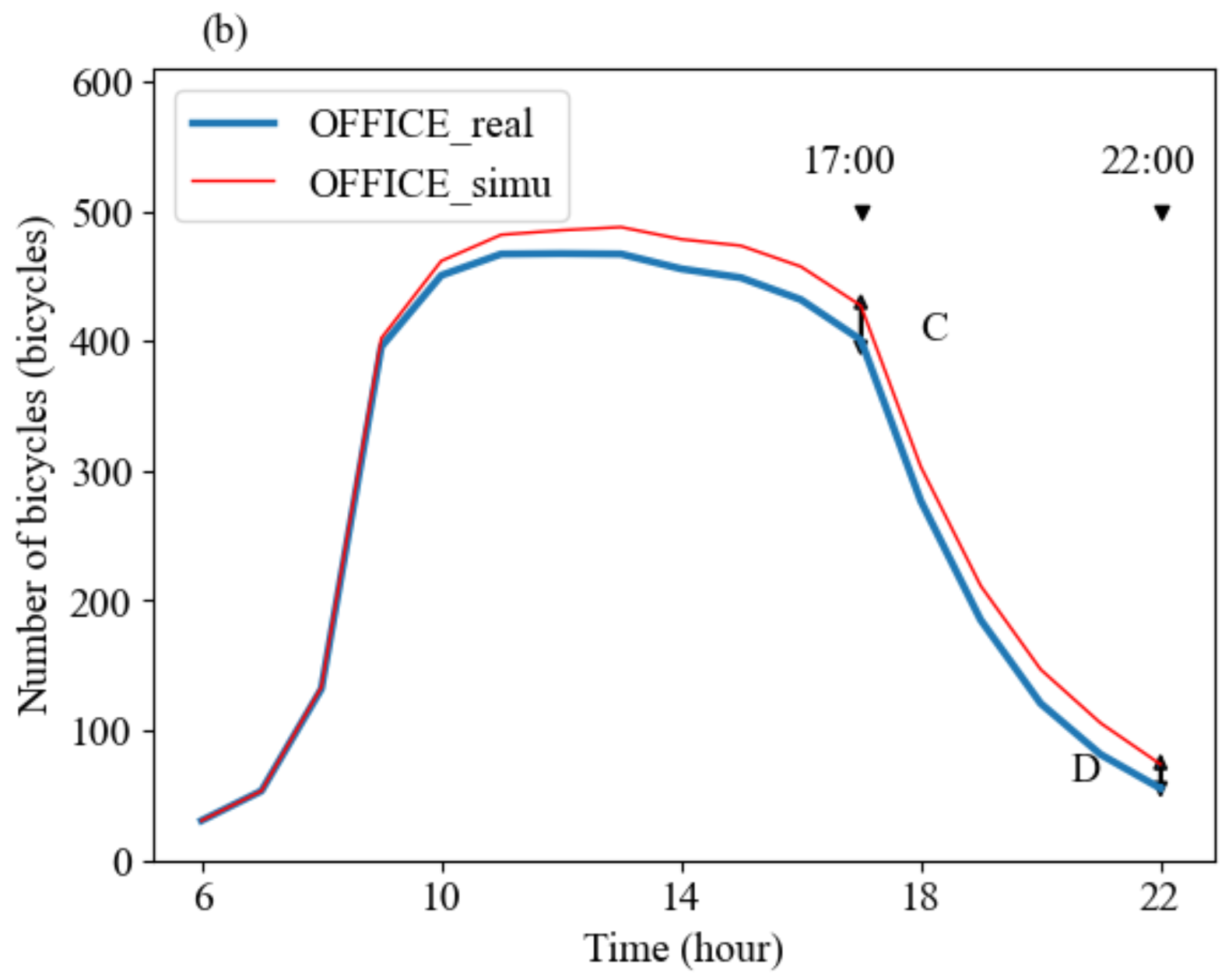}
    \end{minipage}

    \vspace{0.03\linewidth} 
    \begin{minipage}{0.8\linewidth}
        \centering
        \includegraphics[width=\linewidth]{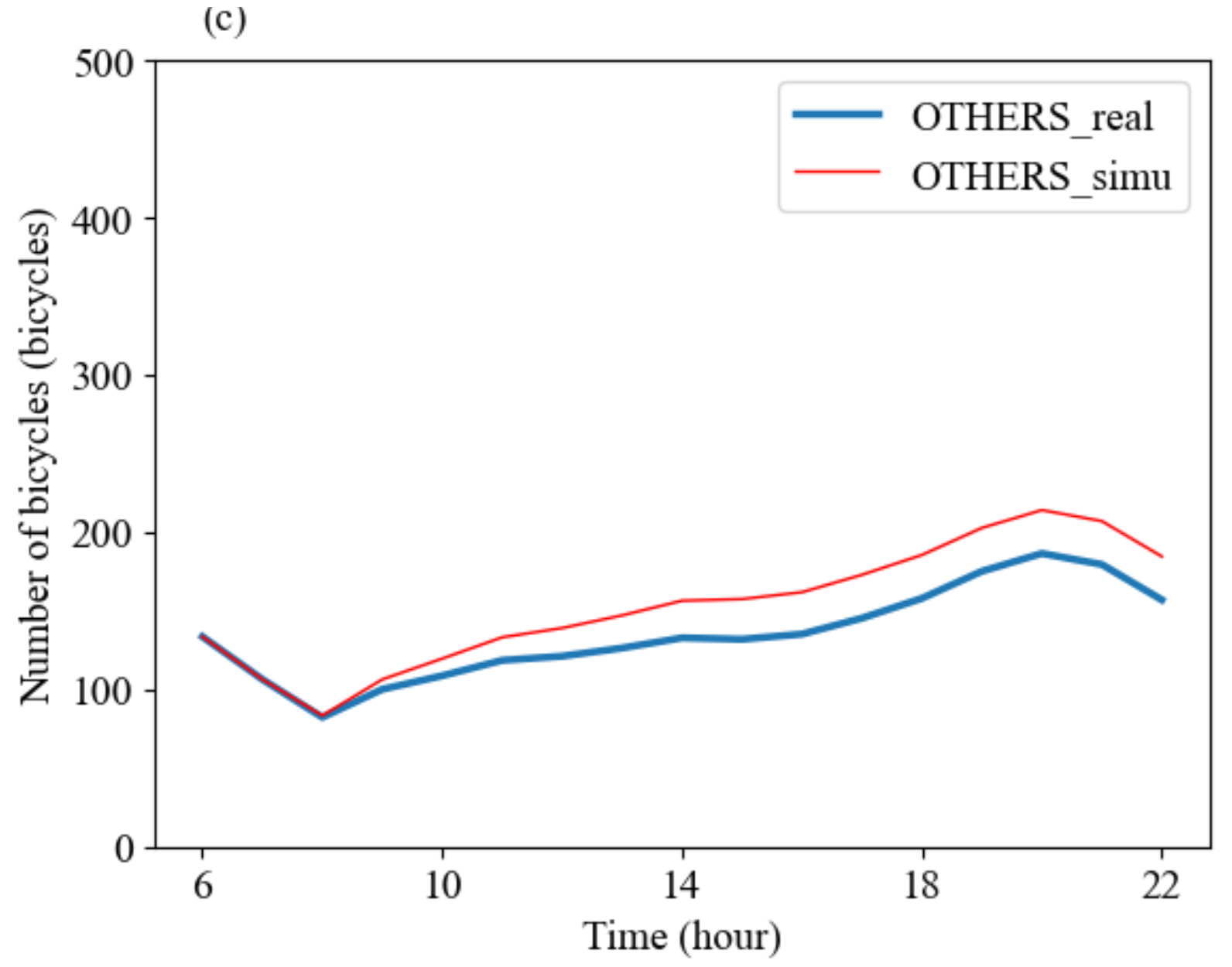}
    \end{minipage}
\caption{
(Color online)~Comparing the bicycle count curve before and after 100 bicycles were added to the residential area in the morning.
(a) depicts the residential area, showing the primary effect measured.
(b) shows the office area, showing the secondary effect measured.
(c) is the others, and no effect was measured.
}
\label{fig:5_3}
\end{figure}

Figure\;\ref{fig:5_3} shows the change in the number of bicycles during a day, comparing the case before and after 100 bicycles are added. The blue (bold) and red lines indicate before and after the addition, respectively. Both are the average values of 1000 samples.

In the residential area where 100 bicycles were added, shown in Fig \ref{fig:5_3} (a), the difference in the number of bicycles before and after the addition is 100 until 8:00~(A), but the difference then decreases, reaching a minimum value 12 at 17:00~(B). This is a first-order effect in which the additional bicycles mitigated bicycle shortages due to commute to work and obtained extra rentals, resulting increase of bicycle rentals by $100 - 12 = 88$. In the office area shown in Fig \ref{fig:5_3} (b), the arrival of additional bicycles from the residential area increased the number of bicycles toward evening, reaching a maximum difference 26 at 17:00~(C). This difference then began to decrease, reaching 18 at 22:00~(D). This decrease of $26 - 18 = 8$ bicycles is a second-order effect in which the additional bicycles arrived from the residential area alleviated bicycle shortages due to commute to home, creating new rentals. In the \textit{Others} area shown in Fig \ref{fig:5_3} (c), a few ports had bicycle shortages during the day. Thus, no secondary effect was observed.
The simulation results are summarized in Table \ref{table:Increase Number}.

\begin{table}
\label{t1}
\begin{center}
\begin{tabular}{lc}
\hline
\multicolumn{1}{c}{Effect Source} & \multicolumn{1}{c}{Number of Rentals} \\
\hline
Primary Effect in \textit{Residential} & 88\\
Secondary Effect in \textit{Office} & 8 \\
\hline
\hline
Total effect & 96 \\
\hline
\end{tabular}
\end{center}
\caption{Increase number of bicycle rentals.}
\label{table:Increase Number}
\end{table}

\section{Discussion}
\label{chap:Discussion}

This study introduced a quantum generative modeling framework tailored to bicycle-sharing demand, bridging discrete, multi-dimensional time series with parameterized quantum dynamics.
Methodologically, we (i) discretized hourly increments via SAX with $N{=}2$ states per dimension (Section~\ref{sec:Time Series Data}),
(ii) constructed probability transition matrices $T_{d,t}(n|m)$ and mapped them to quantum measurement probabilities $P^{\vec{\theta}}_{d,t}(n|m)$ through the evolution $U(\vec{\theta},t)$ (Section~\ref{sec:quantum circuit}),
and (iii) trained the circuit by minimizing the correlation-aware objective in Eq.~(\ref{eq:cost_func}), which couples marginal transition fidelity with inter-port dependence measured by Eqs.~(\ref{eq:actual_corr})--(\ref{eq:quantum_corr}).

Empirically, the learned generator reproduced characteristic daily trends across the three port groups and recovered the qualitative correlation structure (e.g., negative \textit{Residential}~/~\textit{Office} correlation), as evidenced in Fig.~\ref{fig:4_3} and Table~\ref{table:correlation coefficient comparison}.
Crucially, although each dimension uses only two SAX states ($N{=}2$), mapping measurement outcomes $n_{d,t}\in\{0,1\}$ back to representative increments $A_{d,t}(n_{d,t})$ and accumulating over time yielded realistic, system-scale fluctuations.
The model then enabled a counterfactual intervention: estimating the primary and secondary effects of pre-adding bicycles in the morning using Monte Carlo sampling of multi-dimensional trajectories (Fig.~\ref{fig:5_3} and Table~\ref{table:Increase Number}).
These results illustrate how the quantum circuit, trained to match transition behavior and cross-port correlations, can serve as a descriptive model and a decision-support tool for operational policies.

From a modeling standpoint, the decomposition $U(\vec{\theta},t)=V^\dagger(\vec{\theta}_1)W(\vec{\theta}_2 t)V(\vec{\theta}_1)$ balanced expressiveness and computational efficiency:
$V$ introduced entanglement across ports to encode dependencies, while the diagonal $W$ provided time scaling without re-optimizing the entire unitary for different $t$.
The explicit correlation term in Eq.~(\ref{eq:cost_func}) effectively steers learning beyond per-port marginals towards joint structure, essential in mobility systems where flows couple locations.

There are, however, clear avenues for improvement.
First, we validated the approach on a three-group aggregation; extending to finer spatial resolution will require circuit and training refinements (e.g., structured ans\"atze, sparsity in entangling patterns, or hierarchical/state-sharing schemes) to maintain tractability.
Second, while shot-based classical simulation sufficed here, future work should examine robustness under realistic hardware noise and assess hardware-executed workflow variants.
Third, the discretization level $N$ trades fidelity against sample complexity; adaptive or time-of-day-dependent binning may better capture non-stationary regimes without inflating parameters.
Finally, a theoretical characterization of time-series classes that are well-approximated by the proposed diagonal-in-time decomposition would clarify the scope and limitations.

In summary, by aligning discrete time-series encoding with a correlation-aware quantum evolution, the present framework jointly models intra-port dynamics and inter-port dependencies and supports counterfactual analyses relevant to bicycle-sharing operations.
This is a foundation for larger-scale deployments and principled comparisons with classical generative baselines under equalized parameter budgets and correlation-matching criteria.

\section*{Acknowledgement}
\begin{acknowledgment}
In spring 2024, the Graduate School of Information Sciences at Tohoku University held an open online course titled "Quantum Computing for You, 2nd Party!". Many students and working people from all over the country participated in it. The course, which focused on the social implementation of quantum technology, included two periods: "lecture" and "exercise". The exercise period was set to create practical quantum computer applications through group work.
This paper is a further study of group work.
We thank all the staff who conducted this valuable course and the students who studied with us.
We received financial supports by programs for bridging the gap between R\&D and IDeal society (Society 5.0) and Generating Economic and social value (BRIDGE) and Cross-ministerial Strategic Innovation Promotion Program (SIP) from the Cabinet Office (No. 23836436).
\end{acknowledgment}

\appendix

\section{Port Classification}
\label{App:Port Classification}

\begin{figure}[h]
        \centering
        \includegraphics[width=1.0\linewidth]{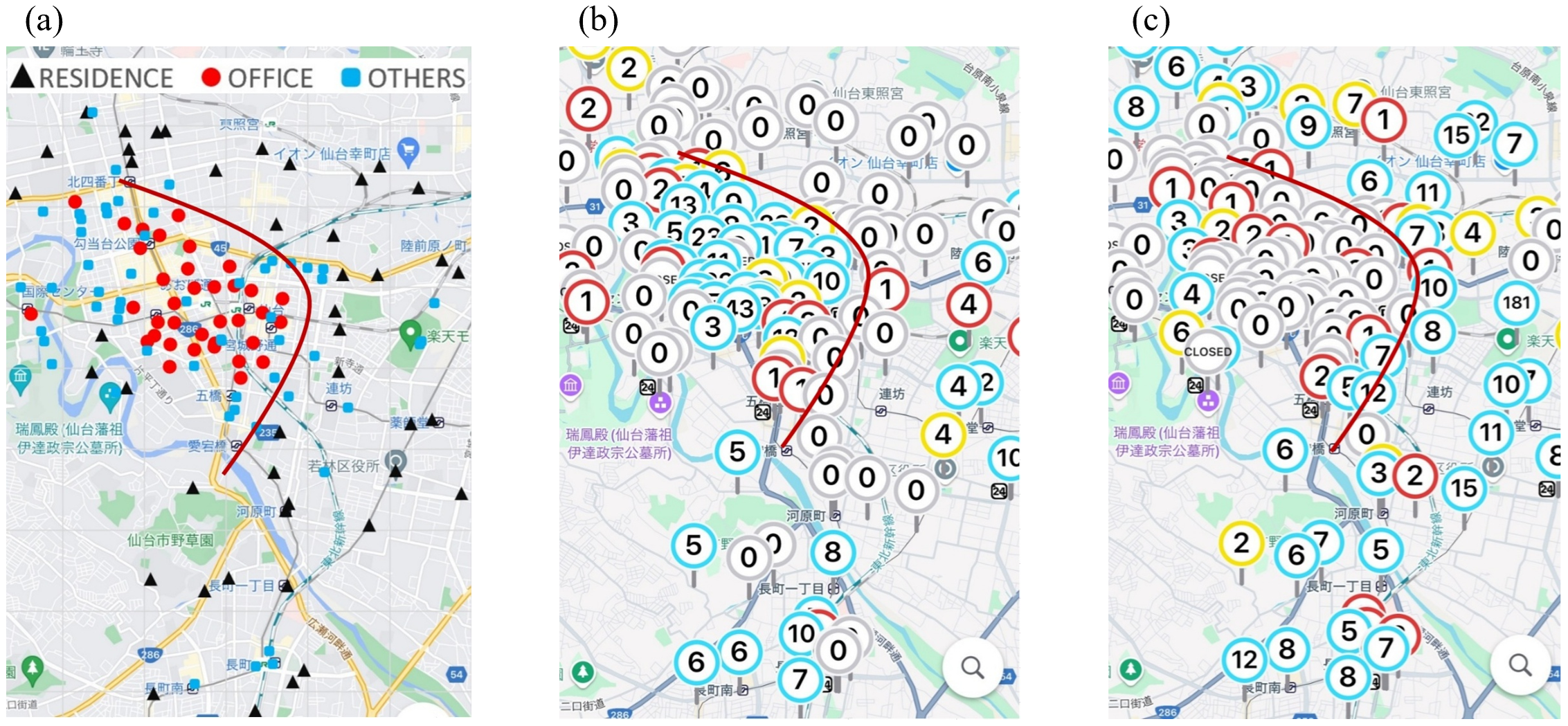}
\caption{(Color online)~(a) Port classification into \textit{Residential}, \textit{Office}, and \textit{Others}, marked with black, red, and blue respectively. The auxiliary curve depicts the boundary between \textit{Residential} and \textit{Office}.
(b) and (c) overwrote (a) with the number of available bicycles at (b)~9:00 and (c)~20:00 on a weekday. The numbers in the circles specify the number of bicycles available.
The colors of circle enclosing indicate the ratio of available bicycles to the number of port racks, with blue being the largest, followed by yellow and red. Ports with zero available bicycles are shown in gray.}
\label{fig:5_1}
\end{figure}

The port classification introduced in Section~\ref{chap:Data Preparation} is consistent with the actual urban structure of Sendai.
Figure~\ref{fig:5_1} (a) shows the classification of DATE BIKE ports overlaid on a city map.
Ports classified as \textit{Office} are concentrated in the area stretching from Sendai Station through Aoba-dori to Kotodai Park, which is widely recognized as the central business district.
These are enclosed within the auxiliary curve in the figure.
Conversely, \textit{Residential} ports include those north of the Kitayonbancho intersection and those along the railway lines, reflecting major residential areas of the city.
These are located outside the auxiliary curve.

(b) and (c) of Fig.~\ref{fig:5_1} are screenshots from the DATE BIKE app at 9:00 and 20:00 on a weekday, respectively.
The numbers in the circles indicate the number of bicycles available at each port.
By comparing these snapshots with the classification map, it is evident that commuter demand is concentrated between \textit{Residential} and \textit{Office} areas: bicycles tend to be depleted in the former during the morning and in the latter during the evening.


\section{Model Stability}
\label{App:Model Stability}

Table \ref{table:stability validation} presents the result of three consecutive trainings of the quantum model to see whether the learned models are stable. The training conditions are the same as in Table \ref{table:correlation coefficient comparison} where $\alpha \equiv 1$, except that the initial parameters were randomly selected. As for MAE,  all trials scored  4 $\sim$ 5, indicating all cases are closely approximate the actual data since the scale of actual data is roughly 100.
In addition, the correlation coefficient of all trials recovered the same characteristics as table \ref{table:correlation coefficient comparison}, negative correlation of \textit{Residential}~/~\textit{Office} and \textit{Others}~/~\textit{Residential} at 6:00 - 11:00 and 16:00 - 22:00 and relatively low correlations values in 11:00 - 16:00.
These result shows that the model is stable.

\begin{table}
\centering
\begin{tabular}{l|cccc}
\hline
 & \multicolumn{3}{c}{model~(${\alpha \equiv 1}$)}    & real\\
 & 1st & 2nd    & 3rd    & \\ \hline\hline
MAE & 4.6   & 5.2   & 5.9   &  \text{-} \\ \hline
Correlation Recov & & & &\\
~~~~6:00-11:00 & & & &\\
 ~~~~~~~~Res / Off & -0.16 & -0.35 & -0.26 & -0.38 \\
 ~~~~~~~~Off / Oth & 0.05  & 0.04  & 0.06  & 0.13 \\
 ~~~~~~~~Oth / Res & 0.02& -0.01  &-0.05  & -0.03 \\
~~~~11:00-16:00 & & & &\\
 ~~~~~~~~Res / Off & -0.14& -0.11  &-0.13  & -0.10 \\
 ~~~~~~~~Off / Oth & -0.09 & 0.00  & -0.03  & -0.11 \\
 ~~~~~~~~Oth / Res & 0.05 & -0.08  & 0.03  & -0.02 \\
~~~~16:00-22:00 & & & &\\
 ~~~~~~~~Res / Off & -0.28 & -0.28  & -0.31  & -0.44 \\
 ~~~~~~~~Off / Oth & -0.09 & -0.07  & 0.03  & -0.02 \\
 ~~~~~~~~Oth / Res & -0.32 & -0.38  & -0.36  & -0.49 \\ \hline

\end{tabular}

\caption{Verification of model stability by three consecutively learned model from "1st" to "3rd" by measuring MAE and correlation coefficients.}
\label{table:stability validation}
\end{table}

\section{Estimating Opportunity Losses}
\label{App:Estimating Opportunity Losses}

\begin{figure}
        \centering
        \includegraphics[width=0.9\linewidth]{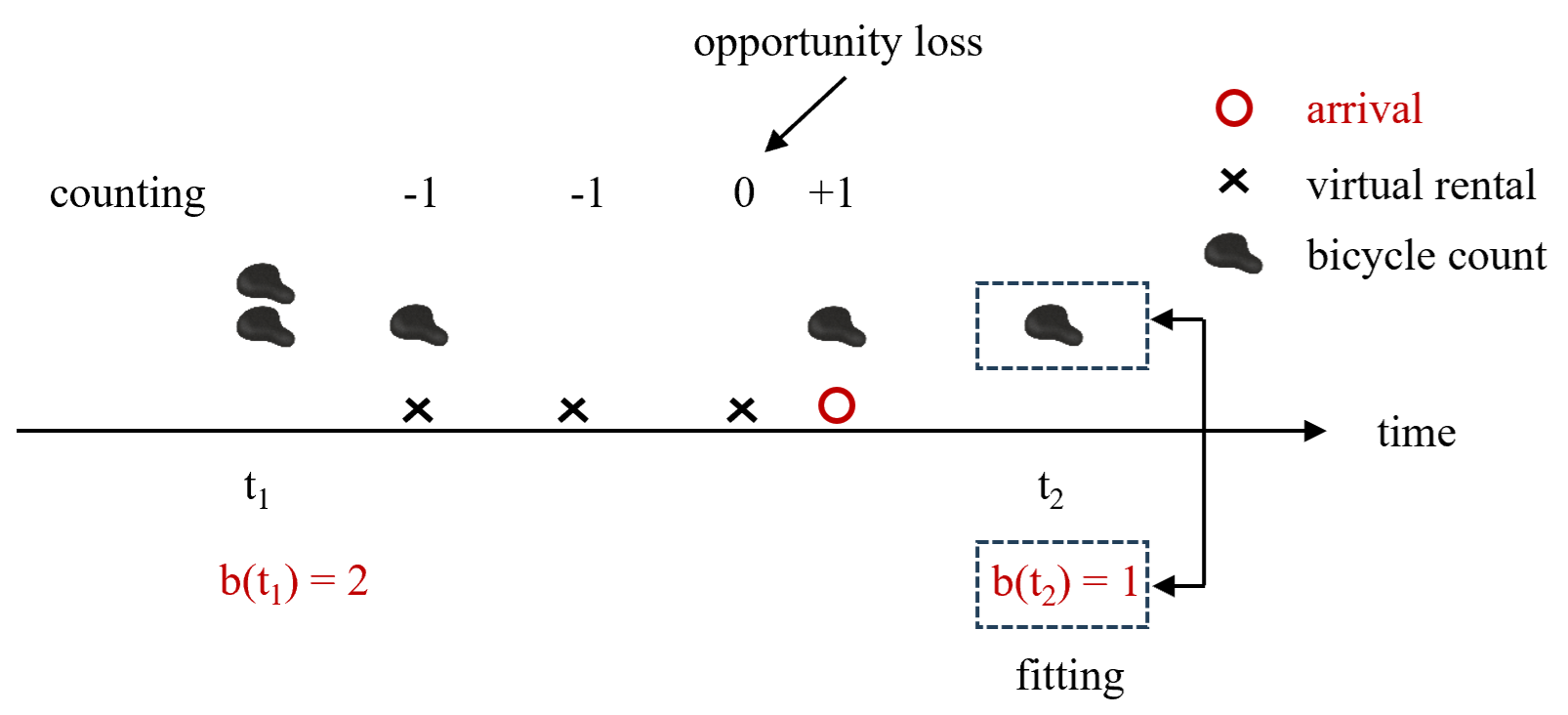}
    \caption{
(Color online)~Illustration of bicycle counts between times $t_1$ and $t_2$.
Red text indicates observed values, including $b(t_1)$, $b(t_2)$, and the arrival sequence $\{t^a_i\}$.
The virtual rentals $\{t^r_i\}$ are estimated so that the simulated count $c(t_2)$ matches the observed $b(t_2)$.}
\label{fig:count model}
\end{figure}

Opportunity losses are defined as the difference between virtual rentals (i.e., rentals that would have occurred if sufficient bicycles had been available) and actual rentals.
We estimate the number of virtual rentals at each port by modeling the fluctuation of bicycle counts, as illustrated in Fig.~\ref{fig:count model}.

We consider the transition of bicycle numbers at a given port between time $t_1$ and $t_2$.
Let $b(t_1)$ and $b(t_2)$ denote the observed number of bicycles at $t_1$ and $t_2$, respectively.
Let $\{t^a_i\}$ be the observed arrival times and $\{t^r_i\}$ the (to-be-estimated) virtual rental times.
By merging $\{t^a_i\}$ and $\{t^r_i\}$ into a single chronological sequence, and starting from $b(t_1)$, we simulate the bicycle count by adding one at each arrival and subtracting one at each rental.
If a virtual rental occurs when no bicycles are available, it is regarded as an opportunity loss and is not counted.
Denoting the simulated count at $t_2$ as $c(t_2)$, we choose the number of elements in $\{t^r_i\}$ so that $c(t_2)$ approximates $b(t_2)$.
Applying this procedure sequentially for each unit interval throughout the day reconstructs bicycle counts under virtual rental conditions.

\begin{figure}
    \centering
    \includegraphics[width=0.9\linewidth]{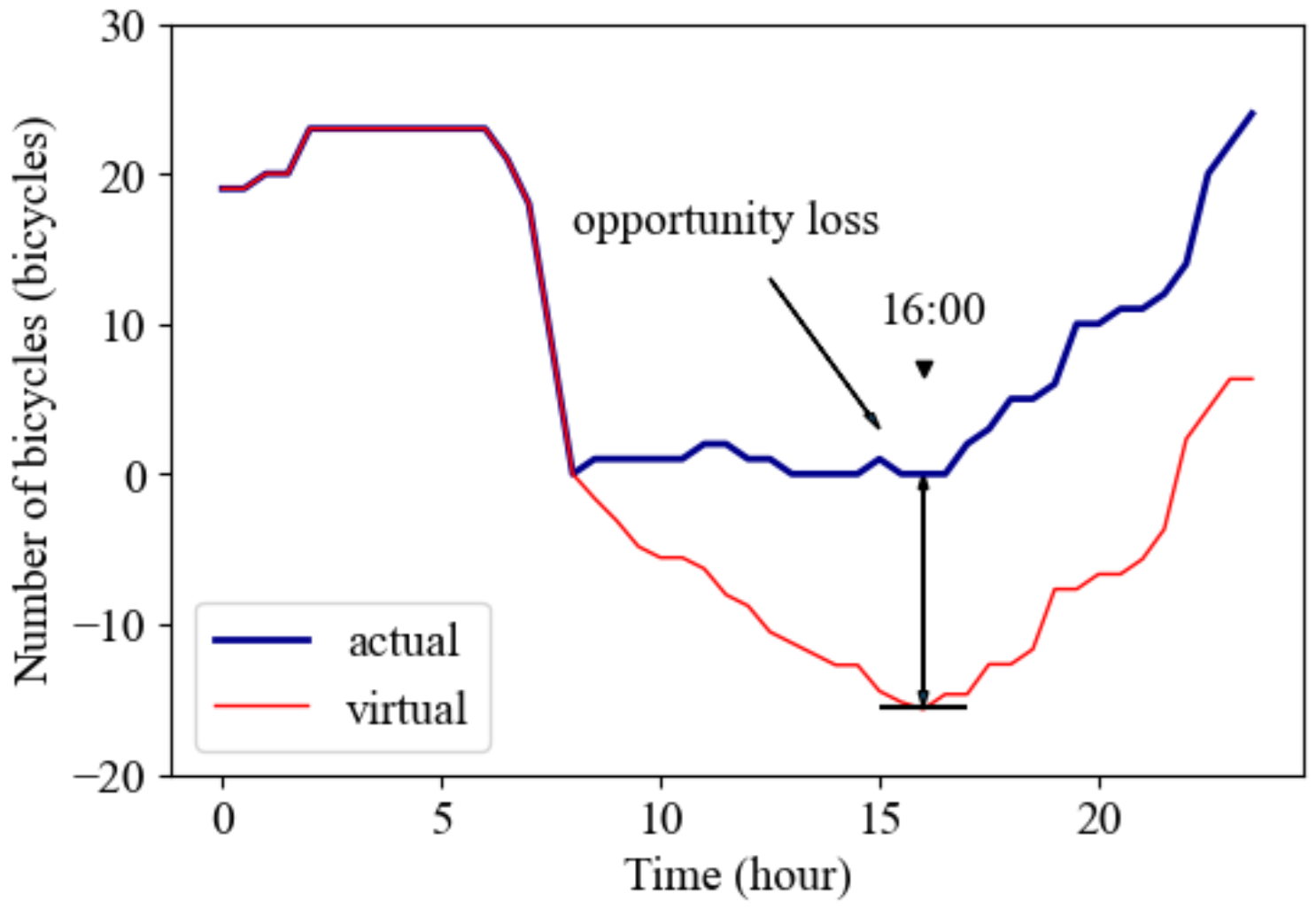}
\caption{(Color online)~Comparison of actual and virtual bicycle numbers at the Kita-Sendai Station port on a weekday.
The black curve shows the actual trajectory, while the red curve represents the trajectory with virtual rentals.
The gap between the two curves corresponds to the number of rental opportunity losses accumulated over time.}
\label{fig:bike count based on virtual rental}
\end{figure}


Figure~\ref{fig:bike count based on virtual rental} demonstrates the estimation for the Kita-Sendai Station port in the \textit{Residential} group.
In the actual trajectory, bicycles are depleted after the morning commute and remain unavailable until around 16:00.
In contrast, the virtual trajectory dips below zero to approximately $-15$, suggesting that 15 additional rentals could have occurred if bicycles had been available.
This gap quantifies the opportunity loss for that day at this port.

\bibliographystyle{jpsj}
\bibliography{shared_bicycle.bib} 
\end{document}